# Physics-Informed Neural Network-based Reliability Analysis of Buried Pipelines

Pouya Taraghi[a], Yong Li[a*], Samer Adeeb[a]

[a] *Department of Civil & Environmental Engineering, University of Alberta, Edmonton, Alberta, T6G 1H9, Canada*
[*]*Corresponding Author's e-mail: yong9@ualberta.ca*

**Abstract:**

Buried pipelines transporting oil and gas across geohazard-prone regions are exposed to potential ground movement, leading to the risk of significant strain demand and structural failure. Reliability analysis, which determines the probability of failure after accounting for pertinent uncertainties, is essential for ensuring the safety of pipeline systems. However, traditional reliability analysis methods involving computationally intensive numerical models, such as finite element simulations of pipeline subjected to ground movement, have limited applications; this is partly because stochastic sampling approaches require repeated simulations over a large number of samples for the uncertain variables when estimating low probabilities. This study introduces Physics-Informed Neural Network for Reliability Analysis (PINN-RA) for buried pipelines subjected to ground movement, which integrates PINN-based surrogate model with Monte Carlo Simulation (MCS) to achieve efficient reliability assessment. To enable its application under uncertain variables associated with soil properties and ground movement, the PINN-based surrogate model is extended to solve a parametric differential equation system, namely the governing equation of pipelines embedded in soil with different properties. The findings demonstrate that PINN-RA significantly reduces the computational effort required and thus accelerates reliability analysis. By eliminating the need for repetitive numerical evaluations of pipeline subjected to permanent ground movement, the proposed approach provides an efficient and scalable tool for pipeline reliability assessment, enabling rapid decision-making in geohazard-prone regions.

# 1. Introduction

Buried energy pipelines are essential for the long-distance transport of oil, gas, and petroleum derivatives. These pipelines often provide a safe, efficient, and reliable means of moving energy resources across vast distances, supporting both domestic and international energy needs. However, they at times traverse regions susceptible to geohazards, such as areas with seismic activity, landslides, liquefaction, and slope instability, where ground movements are prone to occur. Buried continuous pipelines are particularly vulnerable to large-strain related failure (e.g., tensile rupture and buckling), as a result of Permanent Ground Deformations (PGDs). As such, geohazard-induced PGDs present significant risk to pipeline integrity, highlighting the need for strain-based assessments, in which efficient prediction of the strain demand is crucial, particularly in the context of pipeline risk assessment subjected to ground movement.

The risk of pipelines subjected to ground movement is significantly influenced by uncertainties, particularly those in soil properties (e.g., cohesion, internal friction angle, and unit weight), along with the unpredictable nature of geohazard events (e.g., ground movement magnitude). These uncertainties necessitate a probabilistic strain-based assessment, because deterministic methods cannot capture the variation of pipeline response under such uncertainties. As such, reliability and risk analysis are essential for quantifying vulnerabilities within pipeline systems under permanent ground movement, ensuring safe and reliable operation [1–4]. Two primary methods, i.e., semi-quantitative and quantitative, are commonly used in the reliability and risk assessment of buried pipelines. Semi-quantitative methods for assessing pipeline failure risks due to ground movement are associated with significant limitations and challenges. These methods rely on expert judgment and simplified scoring systems, often resulting in inconsistent outcomes and affecting the reliability of the results [5–7].

In contrast, quantitative methods provide a more reliable framework for assessing pipeline failure risks due to ground movement by leveraging numerical simulations and probabilistic analysis. Approximate analytical approaches for structural reliability analysis, such as the First-Order Reliability Method (FORM) and Second-Order Reliability Method (SORM), can be used for this purpose. However, their performance is significantly influenced by the nonlinear behavior of the limit state function, which defines the boundary between safe and failure states. Furthermore, those approaches require gradient information of the implicit limit state function defined based on finite element simulation. These limitations have brought engineers' attention to stochastic simulation approaches, such as Monte Carlo Simulation (MCS). MCS offers a robust and flexible approach that avoids simplifications in the limit state function and eliminates the need for gradient information. However, MCS requires a large number of sample evaluations to achieve reliable estimation of the probability of failure, making it computationally expensive or impractical for real-world engineering applications where evaluating limit state functions involves time-consuming numerical simulations, such as Finite Element Analysis (FEA).

To address the challenge of applying MCS for problems with computationally expensive limit state functions, several alternatives have been proposed to reduce the computational costs of MCS in the probabilistic assessment of pipelines subjected to ground movement. Dey et al. [8] employed a multi-fidelity approach, while Zheng et al. [9] and Phan et al. [10] introduced machine learning-based approaches as alternatives to traditional FEA for limit state function evaluation in the reliability assessment of continuous buried pipelines subjected to ground movement. In another study, Zheng et al. [11] developed finite difference method (FDM) model for buried pipelines subjected to ground movement for its integration with MCS. The computational efficiency of these simulation-based methods, however, still remains challenging because of the number of simulation samples required as the training data for the surrogate model development before MCS or during MCS.

To overcome the limitations associated with stochastic simulation-based methods, this study proposes an efficient approach, by integrating the Physics Informed Neural Networks (PINNs)-based surrogate model with MCS method, for the probabilistic assessment of buried pipelines subjected to ground movement. This approach is referred to as Physics-Informed Neural Network for Reliability Analysis (PINN-RA) of buried pipelines subjected to ground movement. PINNs integrate the governing physics of the problem, encoded as differential equations along with relevant initial and boundary conditions, directly into the neural network's loss function. This integration effectively regularizes the training process, enabling the model to accurately capture the system's behavior without the need for extensive or any training data. In contrast to traditional numerical methods, which require domain discretization and iterative solution procedures, PINNs leverage the mesh-free universal approximation capabilities of neural networks to represent the solution of differential equations, making them particularly effective. Chakraborty [12] pioneered the application of PINNs in reliability analysis by directly incorporating uncertain variables into the surrogate model and transforming the governing equations of the system into parametric differential equations. These equations are solved within the PINN-based approach, providing a universal solution across the defined domain, and the subsequent integration of MCS enables efficient sample evaluation. Following the work of Chakraborty [12], several studies [13–16] have adopted PINNs for uncertainty analysis of various engineering systems.

Different from traditional Machine Learning (ML) approaches for predictive models of pipelines [17–21], the PINN-based surrogate model eliminates the need for any training dataset and overcomes the limitations of traditional ML methods in handling limited or noisy data for real-world applications, establishing PINNs as a computationally efficient surrogate model for uncertainty quantification in pipeline engineering. Taraghi et al. [22–24] pioneered the application of a PINN-based prediction model for pipeline analysis, assuming uniaxial material behavior and incorporating ground movement magnitude as input variable. However, this research extends the PINN-based model by considering the biaxial stress state of steel material while accounting for the effects of temperature and internal pressure. Furthermore, the extended model incorporates a broader range of input variables, which allows to use the PINN-based model directly to capture uncertainties associated with soil properties and ground movement magnitude when using MCS for comprehensive uncertainty analysis of pipelines subjected to PGDs. The proposed model provides the pipeline industry with an efficient tool for conducting reliability analysis, which can be incorporated into a comprehensive risk management program to support the safe operation of pipelines in regions susceptible to geological hazards.

## 2. Methodology

This section presents the proposed PINN-RA approach for buried pipelines subjected to ground movements, which integrates PINN-based surrogate model with MCS to enable efficient reliability assessment. It starts with (1) governing differential equations of buried pipelines subjected to ground movements after accounting for the bi-axial steel material models to consider the effect of internal pressure and temperature change, followed by (2) PINN-based surrogate model development for buried pipelines subjected to ground movements, and (3) the integration of the PINN-based surrogate model with MCS for efficient uncertainty analysis.

*2.1. Governing Differential Equations of Buried Pipelines subjected to Ground Movements*

Figure 1 schematically depicts buried pipelines subjected to ground movements caused by various geohazards, including landslides, seismic faulting, subsidence, and frost heave. The ground movement, characterized by a ground displacement pattern (e.g., rectangular), a magnitude of $\delta$, and an inclination angle $\beta$ relative to the pipe's axial direction, imposes large deformations and high strain demands, posing critical risks to the structural integrity and reliable operation of pipeline systems. To analyze the response of buried pipelines subjected to ground movement, a practical and efficient approach assumes the pipeline as a Euler-Bernoulli beam, with pipe-soil interaction represented by discrete soil springs distributed along its length in the axial and lateral directions, as shown in Fig. (2a). In Figure (2a), $u(x)$ and $w(x)$ refer to the axial and lateral displacements of the centroid of the cross-section $C$ along the pipe's length respectively.

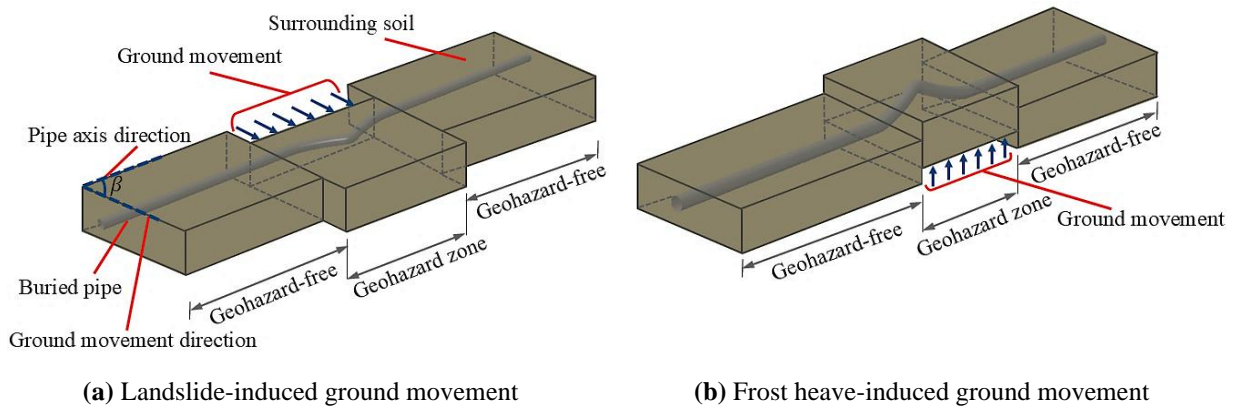

**(a)** Landslide-induced ground movement     **(b)** Frost heave-induced ground movement

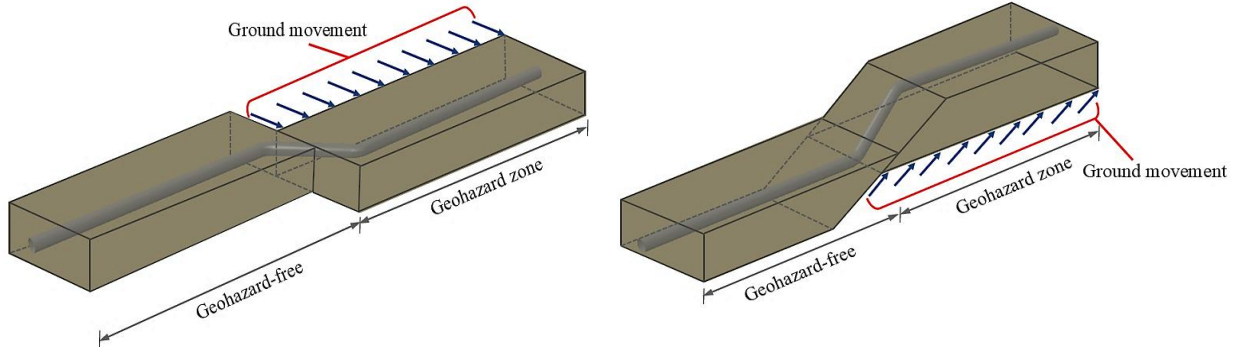

**(c)** Strike-slip fault-induced ground movement      **(d)** Normal fault-induced ground movement

**Fig. 1.** Schematic illustrations of geohazard-induced ground movements affecting buried pipelines

A Euler-Bernoulli beam subjected to ground movement, is described by Eq. (1), following the equilibrium principle of an infinitesimal beam segment after incorporating moderately large deformations to capture geometric nonlinearity. The axial and lateral soil forces per unit length, denoted as $h$ and $q$, respectively, are determined by the relative displacements between the pipeline and the applied ground movement. Specifically, the axial load density ($h$) is a function of the difference between the axial displacement of the pipe, $u(x)$, and the axial component of the ground movement, $U_g(x)$, while the lateral load density ($q$) is a function of the difference between the lateral displacement of the pipe, $w(x)$, and the lateral component of the ground movement, $W_g(x)$. Note that, in Eq. (1), $N(x)$ and $M(x)$, represent the internal axial force and bending moment in the pipe, respectively, while $N_x(x)$ and $M_{xx}(x)$ denote the first- and second-order derivatives of $N(x)$ and $M(x)$ with respect to $x$, where $x$ refers to the position along the pipe's length. Furthermore, $\frac{d(\cdot)}{dx}$ denotes the first-order differential operator with respect to $x$. The internal axial force $N(x)$ and bending moment $M(x)$ are resulted from the normal stresses over the pipe's cross section arising from various sources, including ground movement, internal pressure, and temperature variations (see Fig. 2b).

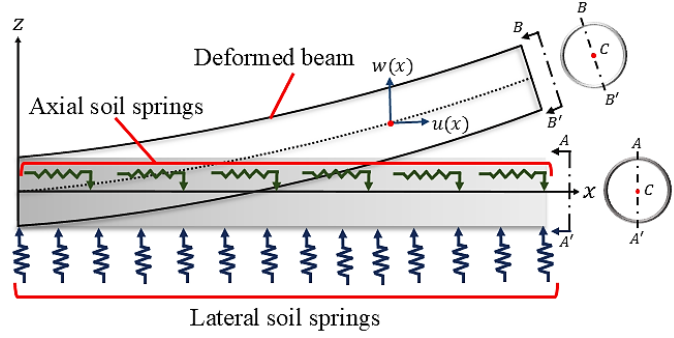

(a) Euler-Bernoulli beam with soil springs representing pipe-soil interaction

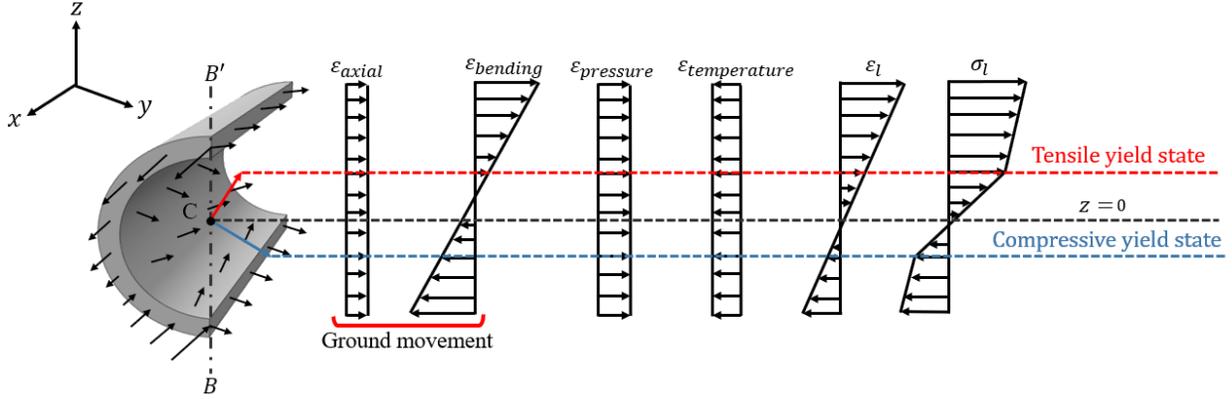

(b) Stress/strain distributions across pipe's cross-section

**Fig. 2.** Buried pipeline represented by the Euler-Bernoulli beam with discrete soil springs

$$\begin{cases} N_x(x) + h\left(U_g(x) - u(x)\right) = 0 & (1a) \\ M_{xx}(x) - \dfrac{d}{dx}\left(N(x) \times w_x(x)\right) - q\left(W_g(x) - w(x)\right) = 0 & (1b) \end{cases}$$

*2.1.1. Longitudinal Strain in Pipes Due to Ground Movement*

The longitudinal strain induced in buried pipelines due to ground movement, denoted as $\varepsilon_{GM}$, arises from the combined effects of axial and bending deformations. Ground-induced pipe displacement generates axial elongation or compression, referred to as axial strain $\varepsilon_{axial}$, and flexural deformation, characterized by bending strain $\varepsilon_{bending}$, as shown in Eq. (2).

$$\varepsilon_{GM}(x,z) = (\varepsilon_{axial}) + (\varepsilon_{bending}) \cong \left(u_x(x) + \frac{1}{2}w_x^2(x)\right) - (z \times w_{xx}(x)) \qquad (2)$$

Here $u_x$ and $w_x$ denote the first derivatives of the axial and lateral displacements, respectively; $w_{xx}$ corresponds to the second derivative of the lateral displacement; *x* denotes the location of the cross-section along the pipeline and *z* denotes the location along the cross-sectional axis (i.e., BB' in Fig.2b). Note that, the functional dependency on ($x$) and ($z$) is dropped off hereinafter for the sake of conciseness.

*2.1.2. Longitudinal Strain in Pipes Due to Internal Pressure and Temperature Change*

Pipelines subjected to internal pressure experience a biaxial stress state, comprising longitudinal stress in the axial direction and hoop stress in the circumferential direction around the pipe wall as shown in Fig. 3. The longitudinal stress arises along with the hoop stress $\sigma_h$ due to Poisson effect. The corresponding longitudinal strain $\varepsilon_h$ is related to the longitudinal stress through Hooke's law, as shown in Eq. (3), where $P$ denotes the internal pressure, $D$ is the outer diameter of the pipe, $t$ represents the pipe's wall thickness, $\nu$ is the Poisson ratio, and $E$ refers to the Young's modulus of pipe steel.

$$\varepsilon_h = \frac{\sigma_h}{E}, \text{in which } \sigma_h = \nu\frac{P(D-2t)}{2t} \qquad (3)$$

In pipelines subjected to temperature variations, axial strain arises due to constrained thermal expansion or contraction. The thermal strain considered here, expressed in Eq. (4), is purely axial, assuming uniform temperature distribution along the pipeline length for the cases considered in this study.

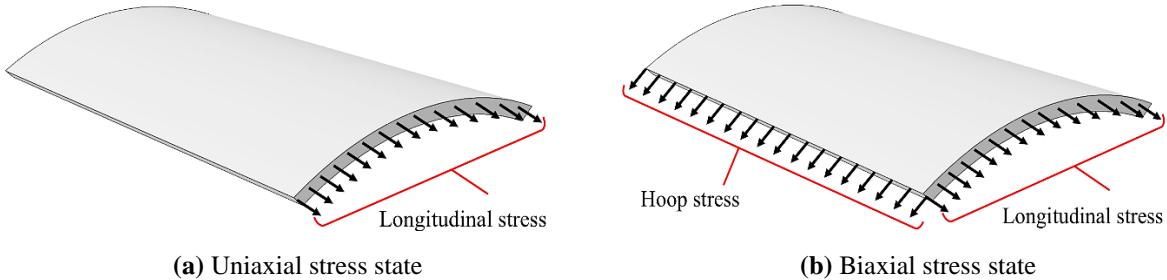

**(a)** Uniaxial stress state  **(b)** Biaxial stress state

**Fig. 3.** Stress states in pipes with and without internal pressure

$$\varepsilon_{\Delta T} = \alpha \Delta T \qquad (4)$$

where $\alpha$ is the thermal expansion coefficient of the pipeline material, and $\Delta T$ is the temperature difference between the installation temperature prior to operation and the temperature at the onset of ground movement (i.e., no temperature gradient is considered). Note that the strain is compressive for increasing temperatures and tensile for decreasing temperatures.

For pipelines subjected to the combined effects of internal pressure and temperature variations, a pseudo-initial strain, $\varepsilon_{initial}$, develops uniformly over the cross-section prior to the onset of ground movement, as defined in Eq. (5).

$$\varepsilon_{initial} = \varepsilon_h + \varepsilon_{\Delta T} \tag{5}$$

Therefore, the total longitudinal strain $\varepsilon_l$ of pipelines under operational loads and geohazard-induced ground movement is formulated in Eq. (6).

$$\varepsilon_l = (\varepsilon_{initial}) + (\varepsilon_{GM}) = \left(\frac{\nu P(D-2t)}{2t} + \alpha \Delta T\right) + \left(u_x + \frac{1}{2}w_x^2 - z \times w_{xx}\right) \tag{6}$$

*2.1.3. Constitutive Equations for Longitudinal Stress-Strain of Pipe under Biaxial Stress State*

The combined effect of the longitudinal stress $\sigma_l$ and hoop stresses $\sigma_h$ in the biaxial stress state within the pipe material can be evaluated using the Von Mises yield criterion based on $\sigma_{vM}$ as defined in Eq. (7). To determine the tensile and compressive yielding stresses for $\sigma_l$, denoted as $\sigma_y^T$ and $\sigma_y^C$ respectively after taking into account the hoop stress effect, Eq. (7) is solved for $\sigma_l$ as expressed in Eq. (8) and Eq. (9), after replacing $\sigma_{vM}$ with the uniaxial yield stress $\sigma_y$ in Eq. (7) [25]. Figure 4 illustrates the stress-strain curves for $\sigma_l$ when the pipe is in the uniaxial and biaxial stress states, with comparison showing a higher tensile yield stress and lower compressive yield stress in the biaxial state due to the tensile hoop stress ($\sigma_h > 0$) compared to the uniaxial state.

$$\sigma_{vM} = \sqrt{\sigma_l^2 + \sigma_h^2 - \sigma_l \sigma_h} \tag{7}$$

$$\sigma_y^T = \frac{1}{2}\left(\sigma_h + \sqrt{4\sigma_y^2 - 3\sigma_h^2}\right) \tag{8}$$

$$\sigma_y^C = \frac{1}{2}\left(\sigma_h - \sqrt{4\sigma_y^2 - 3\sigma_h^2}\right) \tag{9}$$

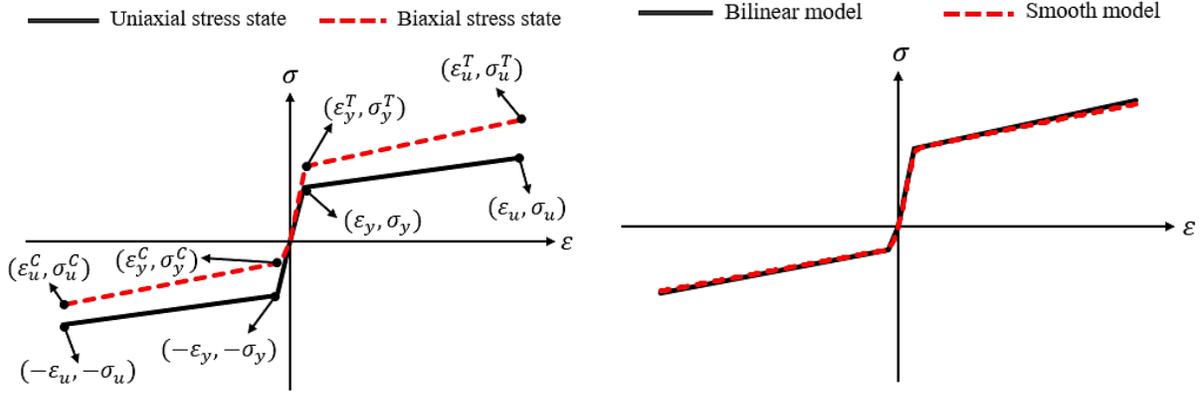

**(a)** Comparison of uniaxial and bi-axial stress states  **(b)** Comparison of bilinear and smooth models

**Fig. 4.** Longitudinal stress-strain behaviors in the axial direction of pipe under uniaxial and bi-axial stress states

The post-yield behavior of the pipe material is characterized using the plasticity flow rule, with an infinitesimal increment of longitudinal strain during plastic deformation derived and expressed mathematically in Eq. (10), in which the parameter $H_p$ is calculated using Eq. (11), with $E$ and $E_p$ referring to the elastic and plastic moduli of the uniaxial stress state, respectively. To calculate the tensile and compressive longitudinal strains in the inelastic region when pipe is under the biaxial stress state, i.e., $\varepsilon_l^T$ and $\varepsilon_l^C$, Eqs. (12) and (13) are obtained by integrating $d\varepsilon_l$ in Eq. (10). Note that, for the sake of simplicity, the longitudinal stress in the plastic region of the biaxial stress state can be approximated as linear with the strain, as shown in Fig. (4a), indicating that the tensile and compressive plastic moduli, i.e., $E_p^T$ and $E_p^C$, can be treated as constant values.

$$d\varepsilon_l = \left(\frac{1}{E} + \frac{(2\sigma_l - \sigma_h)^2}{4H_p(\sigma_l^2 + \sigma_h^2 - \sigma_l\sigma_h)}\right) d\sigma_l \quad (10)$$

$$H_p = \frac{EE_p}{E - E_p} \quad (11)$$

$$\varepsilon_l^T = \frac{\sigma_l}{E} + \frac{1}{4H_p} \int_{\sigma_y^T}^{\sigma_l} \frac{(2\sigma_l - \sigma_h)^2}{(\sigma_l^2 + \sigma_h^2 - \sigma_l\sigma_h)} d\sigma_l \quad (12)$$

$$\varepsilon_l^C = \frac{\sigma_l}{E} + \frac{1}{4H_p} \int_{\sigma_y^C}^{\sigma_l} \frac{(2\sigma_l - \sigma_h)^2}{(\sigma_l^2 + \sigma_h^2 - \sigma_l\sigma_h)} d\sigma_l \quad (13)$$

To facilitate the PINN-based solution of the governing differential equation, longitudinal stress-strain behavior needs to be continuous and smooth for pipe under the biaxial stress state. Thus, this study approximates the longitudinal stress-strain relationship using a smooth alternative, namely the modified Menegotto-Pinto (M-P) material model [26], formulated in Eqs. (14-16) and illustrated in Fig. (4b). Note that, in Eq. (14), parameter $\omega$ represents the scaling factor that is introduced to smoothly connect the tensile and compressive stress-strain curves. Furthermore, in Eqs. (15) and (16), $E$ represents Young's modulus of the pipe material, $b$ denotes the ratio of the plastic modulus to the elastic modulus, and $R$ governs the smoothness of the transition from the elastic to the plastic state, with the superscripts $T$ and $C$ indicating tensile and compressive, respectively.

$$\sigma_l(\varepsilon_l) = \left(\frac{\sigma_l^T(\varepsilon_l) - \sigma_l^C(\varepsilon_l)}{1 + \exp(-\omega \times \varepsilon_l)}\right) + \sigma_l^C(\varepsilon_l) \tag{14}$$

$$\sigma_l^T(\varepsilon_l) = E^T \varepsilon_l \left( b^T + \frac{1 - b^T}{\left(1 + \left(\frac{E^T \varepsilon_l}{\sigma_y^T}\right)^R\right)^{\frac{1}{R}}} \right) \tag{15}$$

$$\sigma_l^C(\varepsilon_l) = E^C \varepsilon_l \left( b^C + \frac{1 - b^C}{\left(1 + \left(\frac{E^C \varepsilon_l}{\sigma_y^C}\right)^R\right)^{\frac{1}{R}}} \right) \tag{16}$$

*2.1.4. Formulation of Internal Axial Force and Bending Moment*

The internal axial force and bending moment are formulated as in Eqs. (17) and (18), respectively; however, there are no closed-form solutions for the integrations. Thus, they are integrated numerically by discretizing the pipe's cross-section into $n_r$ and $n_\theta$ segments along the radial and circumferential directions, respectively, where $\sigma_{ij}$ represents the stress in each patch or area $A_{ij}$ with $i = 1, 2, .., n_r$ and $j = 1, 2, …, n_\theta$. Subsequently, the total internal axial force and bending moment are obtained by summing the contributions of axial forces and bending moments over all patches, as presented in Eqs. (19) and (20), where $z_{ij}$ denotes the vertical distance from the center of patch $A_{ij}$ to the centroid of the cross-section. By substituting the computed internal axial force and bending moment into Eq. (1), the equations are formulated as a system of semi-explicit differential equations in terms of the pipe's axial displacement, $u(x)$, and lateral displacement, $w(x)$.

$$N(x) = \int_A \sigma_l \, dA = \int_A \left( \frac{\sigma_l^T - \sigma_l^C}{1 + \exp(-\omega \times \varepsilon_l)} + \sigma_l^C \right) dA \tag{17}$$

$$M(x) = -\int_A \sigma_l \, z dA = -\int_A \left( \frac{\sigma_l^T - \sigma_l^C}{1 + \exp(-\omega \times \varepsilon_l)} + \sigma_l^C \right) z dA \tag{18}$$

$$N = \sum_{i=1}^{n_\theta} \sum_{j=1}^{n_r} \sigma_{ij} A_{ij} \tag{19}$$

$$M = -\sum_{i=1}^{n_\theta} \sum_{j=1}^{n_r} z_{ij} \sigma_{ij} A_{ij} \tag{20}$$

*2.1.5. Pipe-Soil Interaction Representation*

To explicitly represent pipe-soil interaction forces in the governing equations, a series of nonlinear springs are typically distributed along the pipeline, with each spring providing resistance forces as a function of relative displacement between the soil and pipe, in accordance with the American Lifelines Alliance (ALA) guideline [27]. Following the ALA guidelines, elastic-perfectly plastic models are adopted to describe the axial soil force $h$ and lateral soil force $q$ as functions of the relative displacement between the pipe and soil, which are characterized by the axial resistance $T_u$ and lateral resistance $P_u$, respectively, with their corresponding yield displacements denoted as $\Delta t$ and $\Delta p$. In this study, a continuous and smooth nonlinear function, e.g., *Tanh* function, is further employed to provide a smooth approximation of the axial and lateral soil forces as functions of the relative displacement between the pipe and soil, see Eqs. (21) and (22). This is to facilitate the automatic differentiation required in the PINN. Note that, the axial and lateral soil resistances ($T_u$ and $P_u$) depend on various factors, including the pipe diameter, depth of burial, soil cohesion, unit weight of the soil, interface friction angle, and properties of the pipe's external coating, and details can be found in [27].

$$h\left(U_g(x) - u(x)\right) = T_u \times Tanh\left(U_g(x) - u(x)\right) \qquad (21)$$

$$q\left(W_g(x) - w(x)\right) = P_u \times Tanh\left(W_g(x) - w(x)\right) \qquad (22)$$

With all above, the governing differential equations can be determined, as summarized in Fig. 5. To solve the governing differential equation and obtain the pipeline response, PINN is employed, as discussed in the following section.

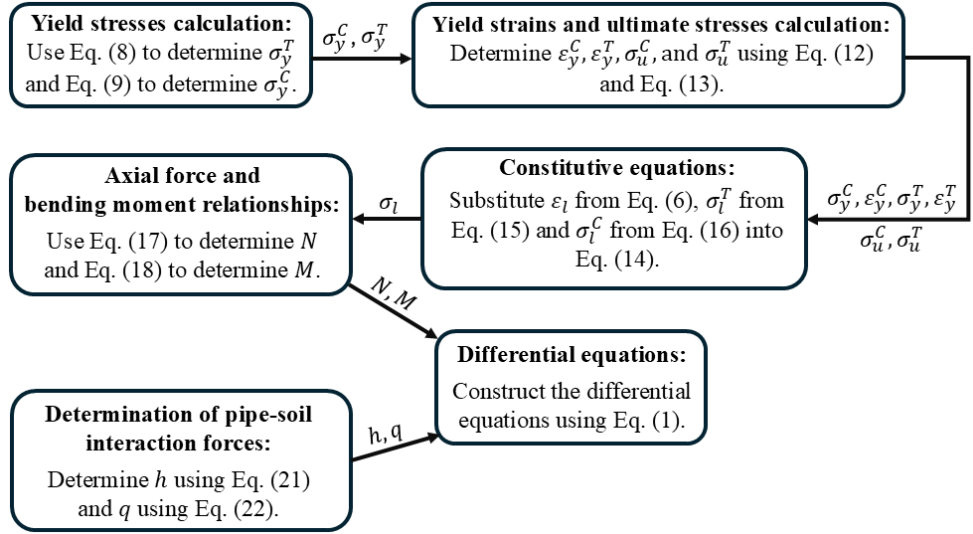

**Fig. 5.** Summary of the formulation of governing differential equations for pipelines subjected to ground movement and operational load effect considering the biaxial stress state

*2.2. PINN-based Surrogate Model for the Responses of Pipeline subjected to Ground Movements*

In this study, Physics-Informed Neural Networks (PINNs), a subclass of feed-forward neural networks, employ Fully Connected Deep Neural Networks (FC-DNNs) with ($L+2$) layers to approximate solutions to governing differential equations for pipelines subjected to ground movement. The approximate solution, $Y$, is computed through a series of nested nonlinear transformations derived from the outputs of successive hidden layers, as expressed in Eq. (23). The $i_{th}$ layer of the network utilizes a nonlinear activation function, $\emptyset_i(\cdot)$, such as ReLU, leaky ReLU, Sigmoid, or hyperbolic tangent, with $z_0$ representing the input of the PINN model (i.e., the first layer), namely, $z_0 = (x; \delta, c, \varphi, \gamma, H)$, and the resulting output $z_i$ ($i = 1, 2, \ldots$) serving as the input to the subsequent layer. The parameters of the $i_{th}$ layer are defined by a weight matrix $W_i$ with dimensions $m \times n$ and a bias vector $b_i$ with dimensions $m \times 1$, where $m$ and $n$ denote the number of nodes in the $i_{th}$ and $(i + 1)_{th}$ layers, respectively. For the sake of brevity, $\theta$ is defined as the collective representation of all network parameters, including the weight matrices and bias vectors across all layers, as described in the Eq. (24), where $NN(\cdot)$ refers to the neural network.

$$\begin{cases} Y = \emptyset_{L+1}(W_{L+1}z_L + b_{L+1}) \\ z_L = \emptyset_L(W_L z_{L-1} + b_L) \\ z_{L-1} = \emptyset_{L-1}(W_{L-1}z_{L-2} + b_{L-1}) \\ \vdots \\ z_1 = \emptyset_1(W_1 z_0 + b_1) \end{cases} \quad (23)$$

$$Y = NN(z_0; \boldsymbol{\theta}) \quad (24)$$

PINNs incorporate the governing physics of the problem, as derived in the previous section and expressed in the parametric form of Eq. (25) without loss of generality, by imposing them as constraints in terms of a penalty term within the loss function in the training process. This set of parametric differential equations will be used to develop the PINN-based surrogate model.

$$\begin{cases} N_x(x; \delta, c, \varphi, \gamma, H) + h\left(U_g(x) - u(x; \delta, c, \varphi, \gamma, H)\right) = 0 & (25a) \\ M_{xx}(x; \delta, c, \varphi, \gamma, H) - \dfrac{d}{dx}\left(N(x) \times w_x(x;, \delta, c, \varphi, \gamma, H)\right) - \\ \qquad q\left(W_g(x) - w(x; \delta, c, \varphi, \gamma, H)\right) = 0 & (25b) \end{cases}$$

Conventional machine learning methods train neural networks by optimizing a loss function that measures the deviation between predicted and observed outcomes, iteratively adjusting parameters to reduce discrepancies and enhance predictive performance. In contrast, PINNs train the model by evaluating the residuals of differential equations at a finite set of randomly generated collocation points within the domain and enforcing boundary conditions at collocation points on the boundary, thereby eliminating reliance on training data. Specifically, Eq. (26a) and Eq. (26b) defines the loss terms associated with Eq. (25a) and Eq. (25b), denoted as $\mathcal{L}_1$ and $\mathcal{L}_2$, respectively, evaluated at the collocation points within the domain, represented by the vector $\{X_f^j\}$ ($j = 1, 2, \dots, N_f$). Here, $\hat{u}$ and $\hat{w}$ represents the NN-approximated axial and lateral displacements of the pipe, while $\widehat{N}$ and $\widehat{M}$ denote the NN-approximated axial force and bending moment, respectively. Furthermore, Eq. (26c) defines the $\mathcal{L}_{BC}$, the loss term corresponding to the boundary conditions of the pipe, which measures the discrepancies between the approximate solution and the exact solution and/or their corresponding derivative components evaluated at collocation points on the boundary, represented by the vector $\{X_u^k\}$ ($k = 1, 2, \dots, N_u$), where $u$ and $w$ represent the exact values for axial and lateral displacements under the boundary conditions. Note that, $\{X_f^j\}$ and $\{X_u^k\}$ denote the collocation point vectors, with dimensions $6 \times N_f$ and $6 \times N_u$, respectively, where the value 6 corresponds to the number of input parameters i.e., $x, \delta, c, \varphi, \gamma$ and $H$. Furthermore, $\mathbb{N}_x^i(\cdot)$ represents the differential operator, where $i = 0, 1, 2, \ldots$ denotes the order of differentiation. Specifically, $i = 0$ corresponds to no differentiation, leaving the terms unchanged, while $i = 1$ refers to the first-order differentiation. PINNs approximate the latent solution as $\hat{u}(x, \delta, c, \varphi, \gamma, H; \boldsymbol{\theta})$ and $\hat{w}(x, \delta, c, \varphi, \gamma, H; \boldsymbol{\theta})$ using FC-DNNs, while their derivatives are computed using the automatic differentiation through the chain rule [28], all parametrized by $\boldsymbol{\theta}$.

$$\begin{cases} \mathcal{L}_1(\boldsymbol{\theta}) = \frac{1}{N_f} \sum_{j=1}^{N_f} \left\| \hat{N}_x\left(X_f^j\right) + h\left(U_g\left(X_f^j\right) - \hat{u}\left(X_f^j\right)\right) \right\|^2 & (26a) \\ \mathcal{L}_2(\boldsymbol{\theta}) = \frac{1}{N_f} \sum_{j=1}^{N_f} \left\| \hat{M}_{xx}\left(X_f^j\right) - \frac{d}{dx}\left(\hat{N}\left(X_f^j\right) \times \hat{w}_x\left(X_f^j\right)\right) - q\left(W_g\left(X_f^j\right) - \hat{w}\left(X_f^j\right)\right) \right\|^2 & (26b) \\ \mathcal{L}_{BC}(\boldsymbol{\theta}) = \frac{1}{N_u} \sum_{i=1}^{N_u} \left\| \mathbb{N}_x^i\left(\hat{u}(X_u^k)\right) - \mathbb{N}_x^i\left(u(X_u^k)\right) \right\|^2 + \left\| \mathbb{N}_x^i\left(\hat{w}(X_u^k)\right) - \mathbb{N}_x^i\left(w(X_u^k)\right) \right\|^2 & (26c) \end{cases}$$

Consequently, the total loss, as defined in Eq. (27), is expressed as the sum of the loss terms corresponding to the governing differential equations and the boundary conditions. As such, the training of the PINN is formulated as an optimization problem described in Eq. (28), where $\boldsymbol{\theta}$ is updated iteratively to minimize the loss function in Eq. (27) and determine the optimal value $\boldsymbol{\theta}^*$. The $NN$ developed based on PINN serves as a multi-input-multi-output (MIMO) surrogate model that maps the inputs $(x; \delta, c, \varphi, \gamma, H)$ to the responses of buried pipelines subjected to ground movements, such as axial displacement $\hat{u}(x; \delta, c, \varphi, \gamma, H)$, lateral displacements $\hat{w}(x; \delta, c, \varphi, \gamma, H)$, and longitudinal strains as a function of $(x; \delta, c, \varphi, \gamma, H)$.

$$\mathcal{L}_{total}(\boldsymbol{\theta}) = \mathcal{L}_1(\boldsymbol{\theta}) + \mathcal{L}_2(\boldsymbol{\theta}) + \mathcal{L}_{BC}(\boldsymbol{\theta}) \tag{27}$$

$$\boldsymbol{\theta}^* = \underset{\boldsymbol{\theta}}{\operatorname{argmin}} \mathcal{L}_{total}(\boldsymbol{\theta}) \tag{28}$$

*2.3. Integration of PINN-based Surrogate Model with MCS*

Reliability analysis is widely employed to quantitatively evaluate structural safety, determining the failure probability $P_f$, as defined in Eq. (29).

$$P_f = P(g(\boldsymbol{\xi}) \leq 0) = \int_{g(\boldsymbol{\xi}) \leq 0} f(\boldsymbol{\xi}) d\boldsymbol{\xi} \tag{29}$$

Here, $g(\cdot)$ represents the limit state function defined such that its negative value indicates failure or undesired performance, $\boldsymbol{\xi}$ denotes the random variable vector describing the uncertainties in the system, and $f(\boldsymbol{\xi})$ refers to the probability density function (PDF) of the random variables. The evaluation of failure probability $P_f$ presents significant challenges for the integration due to the complexity of the limit state function, arising from its implicitness, high nonlinearity, and/or high dimensionality. Nevertheless, the MCS method provides a robust numerical approach to compute $P_f$ through random sampling. Specifically, the MCS method estimates the failure probability by evaluating the limit state function at a large set of $N$ random samples to determine the occurrence rate of failure. The Probability of Failure (PoF) is then estimated as the ratio of failure instances to the total number of simulations $N$, as mathematically expressed in Eq. (30), where $I(\cdot)$ specifies the indicator function that equals one when $g(\boldsymbol{\xi}_i) \leq 0$ and zero otherwise.

$$P_f \approx \frac{1}{N} \sum_{i=1}^{N} I(g(\boldsymbol{\xi}_i) \leq 0) \tag{30}$$

The acceptance criteria for longitudinal strain resulting from ground movement combined with operational loads are defined in ALA guidelines [27], where the tensile strain capacity $\varepsilon_{ST}^c$ is 2%, representing the strain that a girth weld can sustain without a leak or rupture [11,29], and the compressive strain capacity $\varepsilon_{SC}^c$ is defined in Eq. (31), corresponding to the maximum allowable longitudinal compressive strain that the pipe can sustain without experiencing wrinkling, considering both geometric resistance and the pressure-induced stiffening [30,31]. Consequently, the limit state function is formulated in Eq. (32), where $\varepsilon^c$ and $\varepsilon^d(\cdot)$ represent the strain capacity and strain demand predicted by the surrogate model, respectively, to describe the strain demand exceeding the strain capacity.

$$\varepsilon_{SC}^c = 0.50 \left(\frac{t}{D}\right) - 0.0025 + 3000 \left(\frac{PD}{2tE}\right)^2 \tag{31}$$

$$g(\delta, c, \varphi, \gamma, H) = \varepsilon^c - \varepsilon^d(\delta, c, \varphi, \gamma, H) \tag{32}$$

Figure 6 schematically shows the method of Physics Informed Neural Network for Reliability Analysis (PINN-RA) of pipelines subjected to PGDs, considering ground movement magnitude and soil properties as random variables. As shown in Fig. 6, the semi-parametrized governing differential equations described in section 2.2 are solved within the PINN-based approach, leading to a surrogate model where five parameters, including ground movement magnitude $\delta$, soil cohesion $c$, friction angle $\varphi$, unit weight of soil $\gamma$, and burial depth $H$ that characterize pipeline loading, serve as inputs to the FC-DNN. Within the PINN framework, the model approximates the axial and lateral displacements, while the internal axial force and bending moment are computed from displacement derivatives using nonlinear functions and numerical integration, as detailed in Section 2.1.4. The loss function is formulated using the derivatives of the axial force and bending moment, which are computed through automatic differentiation. Unlike conventional studies that apply automatic differentiation only once to the network outputs and directly incorporate the resulting derivatives into the loss function, this study introduces a two-step automatic differentiation process. Initially, the displacement derivatives ($u_x, w_x, w_{xx}$) obtained from the neural network are passed through highly nonlinear functions to compute the internal axial force $N$ and bending moment $M$. Automatic differentiation is then applied a second time to compute their derivatives, namely $N_x$ and $M_{xx}$ which are incorporated into the loss function formulation. Once trained, the PINN-based MIMO surrogate model provides rapid predictions of the axial displacement, lateral displacement, and strain for various combinations of the uncertain input parameters. The PoF is estimated using MCS, in which stochastic samples of random variables are fed into the limit state function defined based on the PINN-based MIMO surrogate model.

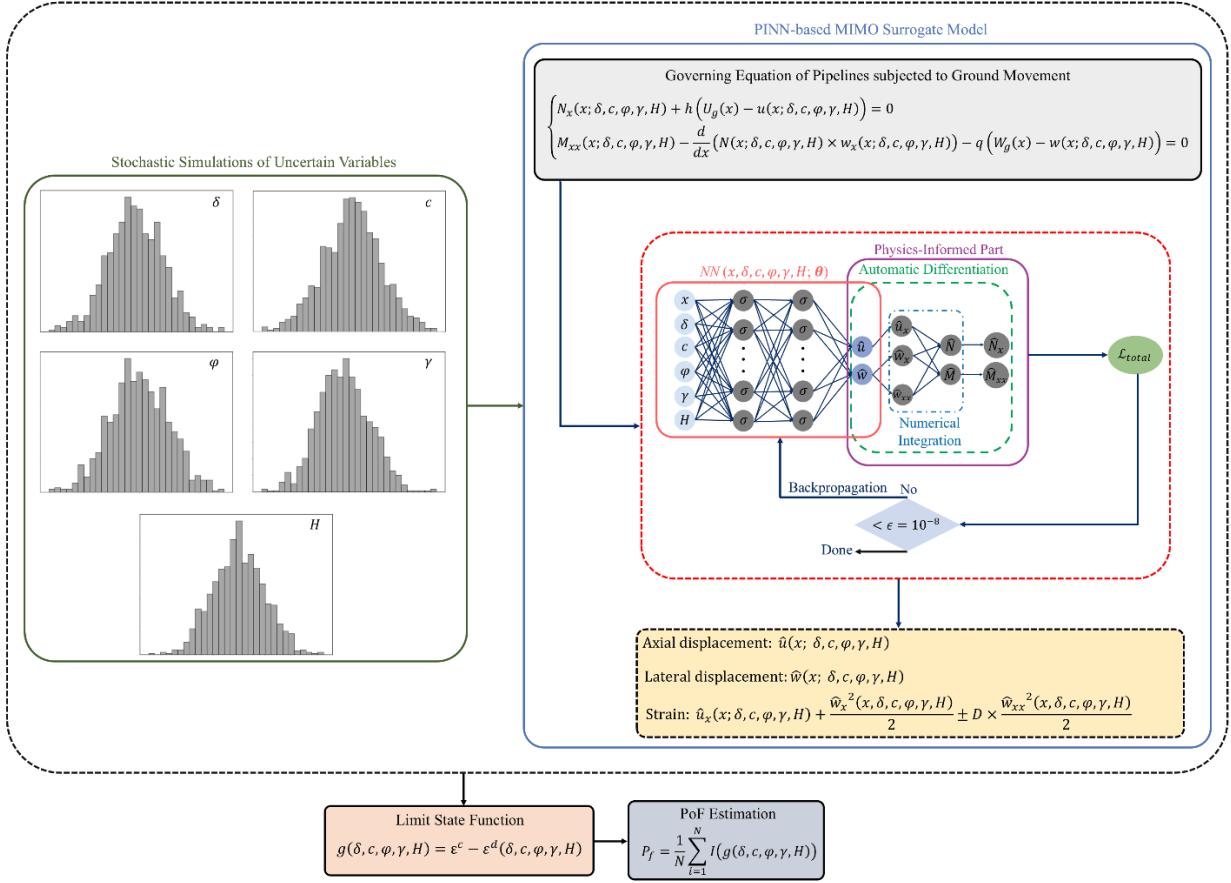

**Fig. 6.** Overview of the PINN-RA method for reliability analysis of pipelines subjected to PGDs, considering ground movement magnitude and soil properties as random variables

## 4. Case Study Simulation of Buried Pipelines Using PINN-Based Surrogate Models

This section uses the developed PINN-based surrogate models for predicting the mechanical response of pipelines subjected to geo-hazard-induced ground movement while considering operational loading conditions, including internal pressure of 10.26 MPa and thermal effects of +60 ℃. To this end, a pipeline with a diameter of 508 mm, a wall thickness of 7.14 mm, and a total length of 90 m is considered. The pipeline is subjected to a geohazard-induced ground movement in a horizontal plane as illustrated in Fig. (7), where a 10 m segment at the center undergoes ground movement with a crossing angle of 90°, while the remaining 40 m on each end is located in a geohazard-free region. To account for the unmodeled length of long-distance pipelines, the affected segment is modeled, and the pipe ends are assumed fixed, restricting axial and lateral deformation and lateral rotation. The pipeline material is API 5L X65 steel, with a Young's modulus of 199 GPa, yield stress of 448 MPa, ultimate stress of 663 MPa, ultimate strain of 0.3, and a coefficient of thermal expansion of $12\times10^{-6}\frac{1}{°C}$. The pipe is buried in soil with varying soil properties and subjected to varying ground movement which are incorporated as input to the PINN-based surrogate model within the range defined in Table 1.

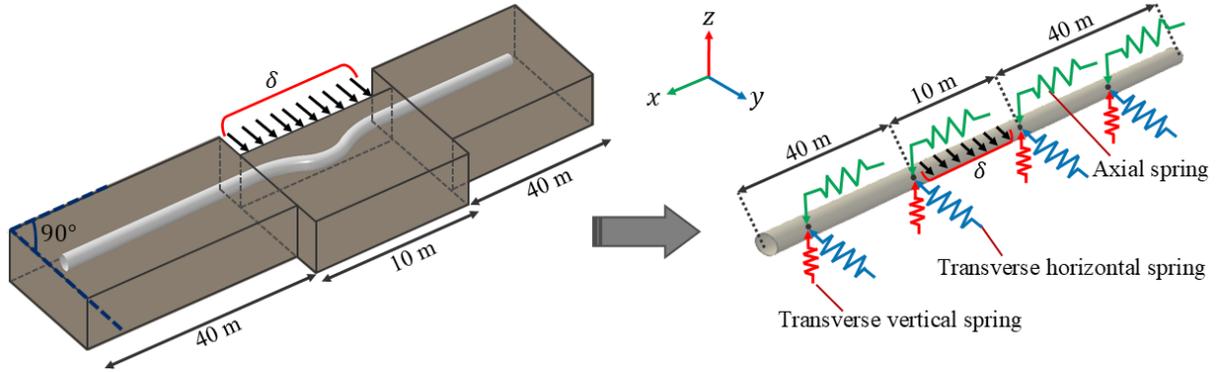

**Fig. 7.** Schematic representation of a pipe subjected to ground movement in the case study

The PINN method is implemented in TensorFlow-GPU (version 1.14.0) [32], an open-source machine learning library optimized for high-performance numerical computation on GPUs, enabling efficient training and inference. The random variables are incorporated as input to the PINN-based surrogate model, which is trained with 20,000 collocation points within the domain and 1,000 collocation points on the boundary, generated randomly using Latin Hypercube Sampling (LHS) within the defined ranges in Table 1. The surrogate model features a fully connected architecture with six hidden layers, each containing 30 neurons ($=\sigma$), and utilizes the hyperbolic tangent as the nonlinear activation function. The L-BFGS-B [33] optimizer is employed to optimize the neural network parameters by minimizing the residual of the loss function, with convergence achieved when the relative error between two consecutive iterations ($=\epsilon$) falls below $10^{-8}$.

**Table 1.** Variations of ground movement and soil properties considered in the PINN-RA model

| Ground movement magnitude, $\delta$ (m) | Soil cohesion, $c$ (kPa) | Friction angle, $\varphi$ (Degree) | Unit weight, $\gamma$ (kN/m³) | Burial depth, $H$ (m) |
|---|---|---|---|---|
| 0.0 − 2.0 | 40 − 50 | 22 − 28 | 15 − 23 | 1.2 − 1.7 |

Figure 8 depicts training loss with respect to the iteration for models under uniaxial and biaxial stress states, where each iteration corresponds to an optimization step for updating the model parameters. The results indicate smooth and rapid convergence, with the loss residual decreasing from $1.37 \times 10^{-1}$ to $3.61 \times 10^{-4}$ over 31,816 iterations for the uniaxial model and from $2.52 \times 10^{-2}$ to $3.72 \times 10^{-4}$ over 34,208 iterations for the biaxial model. Furthermore, Table 2 presents the breakdown of loss function metrics for the training and total test loss, where $\mathcal{L}_1$ and $\mathcal{L}_2$ represent the training loss terms corresponding to the first and second terms of the governing equations, as defined in Eq. (1a) and Eq. (1b), respectively, while $\mathcal{L}_{BC}$ denotes the training loss term associated with the boundary conditions. The results indicate that the residuals of the governing equation loss terms, $\mathcal{L}_1$ and $\mathcal{L}_2$, on the order of $10^{-5}$ and $10^{-4}$, respectively, while the residual of the boundary condition loss term $\mathcal{L}_{BC}$ is on the order of $10^{-6}$, with the total training and test loss residuals on the order of $10^{-4}$. The significantly low residuals of the total loss function and individual loss terms, along with the close alignment between the total training and test loss, indicate that the PINN-RA models effectively converge and learn the solution of the differential equation governing the physics of the problem.

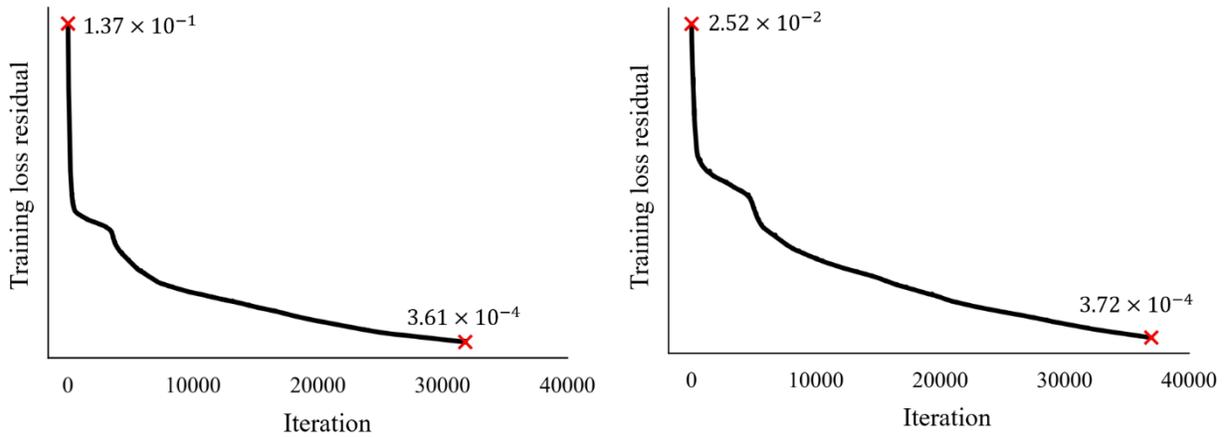

(a) Without internal pressure and temperature effects  (b) With internal pressure and temperature effects

**Fig. 8.** Convergence of the training loss residual for the PINN-RA models: a) Uniaxial stress state, and b) Biaxial stress state

**Table 2.** Loss function metrics for governing equations and boundary conditions

| Case | Governing equations test loss | | Boundary condition test loss | Total loss | |
|---|---|---|---|---|---|
| | $\mathcal{L}_1$ | $\mathcal{L}_2$ | $\mathcal{L}_{BC}$ | Test | Train |
| Uniaxial | $6.10 \times 10^{-5}$ | $3.69 \times 10^{-4}$ | $1.71 \times 10^{-6}$ | $4.32 \times 10^{-4}$ | $3.61 \times 10^{-4}$ |
| Biaxial | $7.96 \times 10^{-5}$ | $3.63 \times 10^{-4}$ | $3.79 \times 10^{-6}$ | $4.46 \times 10^{-4}$ | $3.72 \times 10^{-4}$ |

## 5. Application of the PINN-RA Model for Reliability Analysis of Pipes Under Ground Movement

This section demonstrates the application of PINN-RA approach for reliability analysis of continuous steel pipelines subjected to ground movement. The reliability problem focuses on assessing the probability of excessive strain demand in pipelines when subjected to geohazard-induced ground movements. The pipeline considered earlier is assumed to be subjected to landslide-induced ground movement, with an intersection angle of 90° relative to the pipeline (see Fig. 7). However, the ground movement magnitude is unpredictable with a single value, but the likelihood can be described by a probability model as shown in Table 3. Due to randomness in soil properties such as soil cohesion, friction angle, and unit weight, as well burial depth, they are assumed to follow a normal distribution with statistical properties listed in Table 3 with reference to [5,19,34–37]. The analysis considers two distinct loading conditions, with the pipeline subjected to an internal pressure of 10.26 MPa and a +60°C temperature increase in the first case and operating without internal pressure or temperature loads in the second case.

**Table 3.** Statistical properties of ground movement magnitude and soil properties

| Random variable | Distribution | Mean | Standard deviation |
|---|---|---|---|
| Ground movement magnitude, $\delta$ (m) | Normal | 1.1 | 0.112 |
| Soil cohesion, $c$ (kPa) | Normal | 45 | 0.7150 |
| Friction angle, $\varphi$ (Degree) | Normal | 25 | 0.4260 |
| Unit weight, $\gamma$ (kN/m$^3$) | Normal | 19 | 0.5840 |
| Burial depth, $H$ (m) | Normal | 1.45 | 0.0359 |

*5.1. Analysis and Discussion of Reliability Assessment Using PINN-RA*

To perform an MCS-based reliability analysis using PINN-RA, with soil properties and ground movement magnitude as random variables, the statistical properties summarized in Table 3 are used to generate $10^5$ random samples. To calculate the PoF, two limit state cases are considered including case (I) corresponding to the serviceability limit state defined in Section 2.3 following the ALA guidelines [27] with a tensile strain capacity of 2% and compressive strain capacities of 0.45% and 1.46% under uniaxial and biaxial conditions, respectively, and case (II) introduced for further assessment of the performance of the PINN-RA method with a tensile strain capacity of 2.5% and compressive strain capacities of 0.58% and 1.9% under uniaxial and biaxial conditions, respectively.

The calculated PoF values listed in Table 4 show that for case (I) the PoF is 86.12% and 33.63% for tensile strain and 90.32% and no failure observed for compressive strain, and for case (II) the PoF is 12.07% and 0.001% for tensile strain and 0.024% and no failure observed for compressive strain, in the pipe without and with internal pressure and temperature effects, respectively. Figure 9 illustrates the probability density functions of tensile and compressive strain demand for uniaxial and biaxial models, considering the uncertainties outlined in Table 3. The results reveal that the tensile strain demand exhibits a nearly symmetric distribution with a mean value of approximately 2.25% and a standard deviation of 0.22% for the uniaxial stress state, and a mean of 1.92% with a standard deviation of 0.17% for the biaxial stress state. The relatively wider spread in the uniaxial case reflects greater sensitivity of tensile strain demand to the uncertainties in soil and ground movement parameters, while the reduced dispersion observed under the biaxial condition suggests a more stable tensile response, contributing to improved structural reliability. The compressive strain demand shows a similar bell-shaped distribution in both stress states, centered around −0.50% with a standard deviation of 0.03% for the uniaxial condition and −0.88% with a standard deviation of 0.06% for the biaxial condition. Compared with the tensile strain demand, the compressive strain demand exhibits narrower variability, highlighting that tensile strain is more strongly influenced by the input uncertainties.

**Table 4.** Probability of failure for tensile and compressive strain failure for a pipe with and without internal pressure and temperature effects.

| Case | Without internal pressure and temperature | | With internal pressure and temperature | |
|---|---|---|---|---|
| | Tensile strain failure | Compressive strain failure | Tensile strain failure | Compressive strain failure |
| (I) | 86.12% | 90.32% | 33.63% | No failure |
| (II) | 12.07% | 0.024% | 0.001% | No failure |

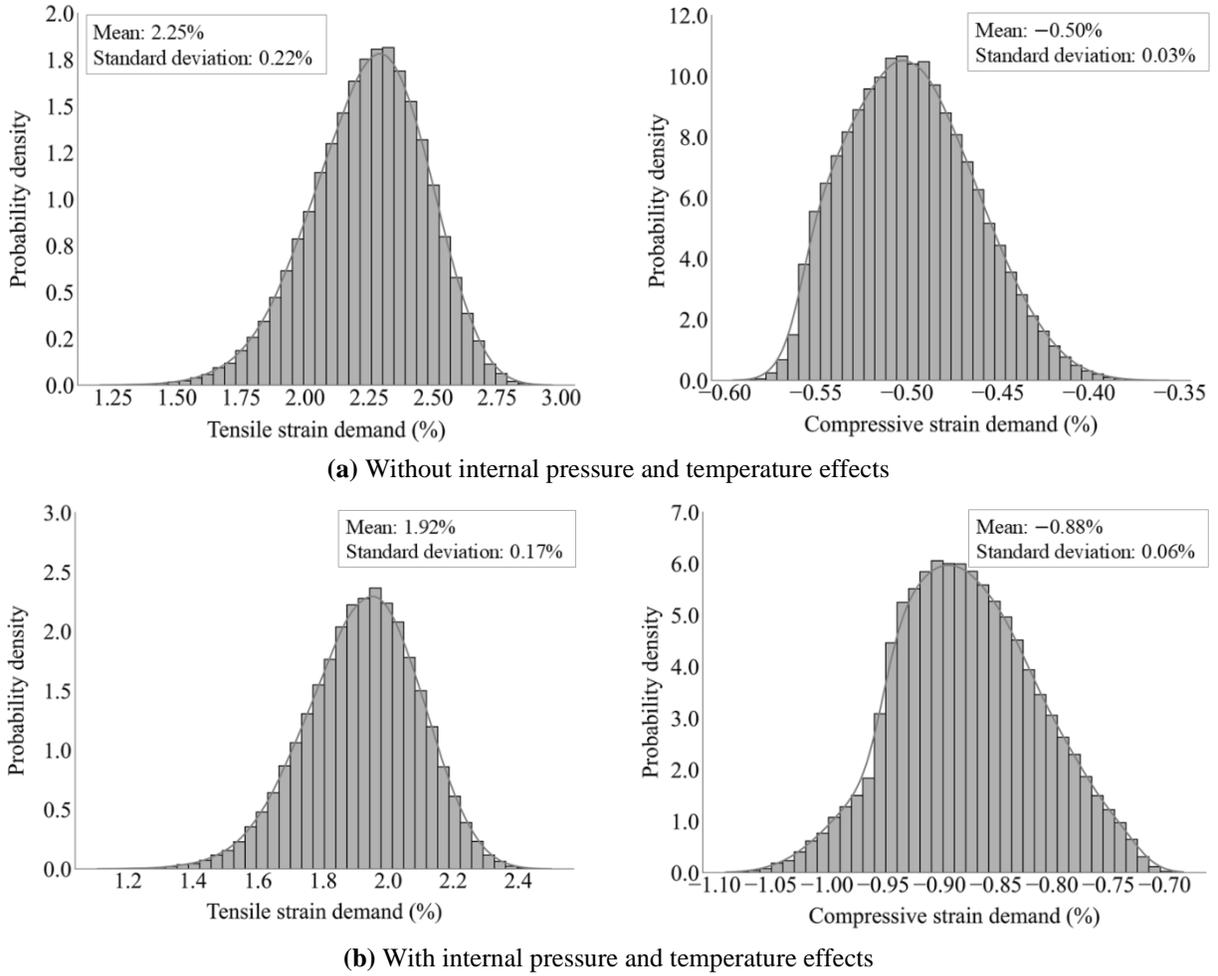

**(a)** Without internal pressure and temperature effects

**(b)** With internal pressure and temperature effects

**Fig. 9.** Probability density functions of tensile and compressive strain demand considering the uncertainties listed in Table 3 under: a) Uniaxial stress state, and b) Biaxial stress state

Sensitivity analysis is conducted to evaluate the influence of input parameter variations on the reliability of buried pipelines and to identify the most critical factors affecting reliability. This is achieved by perturbing each random variable using mean minus standard deviation (Mean - SD) and mean plus standard deviation (Mean + SD) while keeping others constant, allowing for a systematic assessment of their impact. Figure 10 presents the sensitivity index analysis of the reliability assessment for the model with biaxial stress state, with the results highlighting that ground movement magnitude (PGD magnitude) exhibits the highest sensitivity, with index values of 85.57 and 82.85, significantly larger than those of other parameters. This underscores the dominant influence of ground movement uncertainty on the probability of failure, establishing it as the most critical factor in the reliability analysis. Among the soil properties, friction angle, with index values of 12.26 and 12.67, and burial depth, with index values of 11.4 and 7.19, demonstrate a more pronounced influence on reliability variations. In contrast, soil cohesion and unit weight exhibit lower sensitivity, indicating a less significant contribution to reliability assessment.

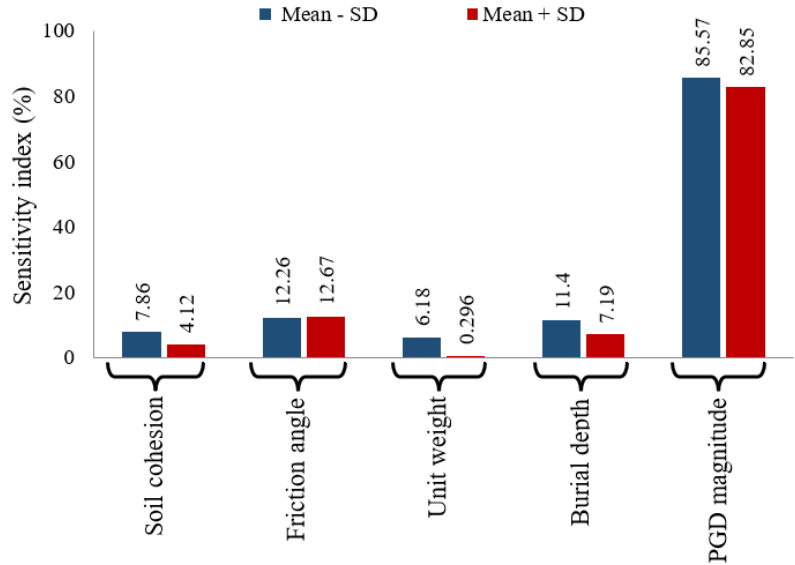

**Fig. 10.** Sensitivity index of reliability for input parameters subjected to $\pm 1$ standard deviation perturbations

Figure 11 presents the computational time required for MCS-based reliability analysis using different methods. The results in Fig. (11a) indicate that evaluating $10^5$ samples with FEM and FDM require 138.9 hours and 55.5 hours, respectively, with each FEM sample taking 5 seconds and each FDM sample taking 2 seconds. Parallelization of FEM with six processors reduces the computation time to 23.2 hours, decreasing the evaluation time per sample to 0.84 seconds. Further optimization, as proposed by Zheng et al. [11], reduces the computational time for FDM to 3.3 hours, with each sample requiring only 0.12 seconds. In contrast, training the PINN-RA model for the biaxial stress state requires 3.1 hours, while sample evaluation takes only a few seconds, remaining independent of the number of samples. To further demonstrate the efficiency of the PINN-RA model, the computational time for performing sensitivity analysis on 10 cases, each with $10^5$ samples, is presented in Figure (11b). The results indicate that the FEM-based model with parallelization using six processors requires 232.2 hours, while the optimized FDM-based model takes 33.3 hours. In contrast, the PINN-RA model requires 3.1 hours for training, with sample evaluation taking only a few seconds, remaining independent of sample size.

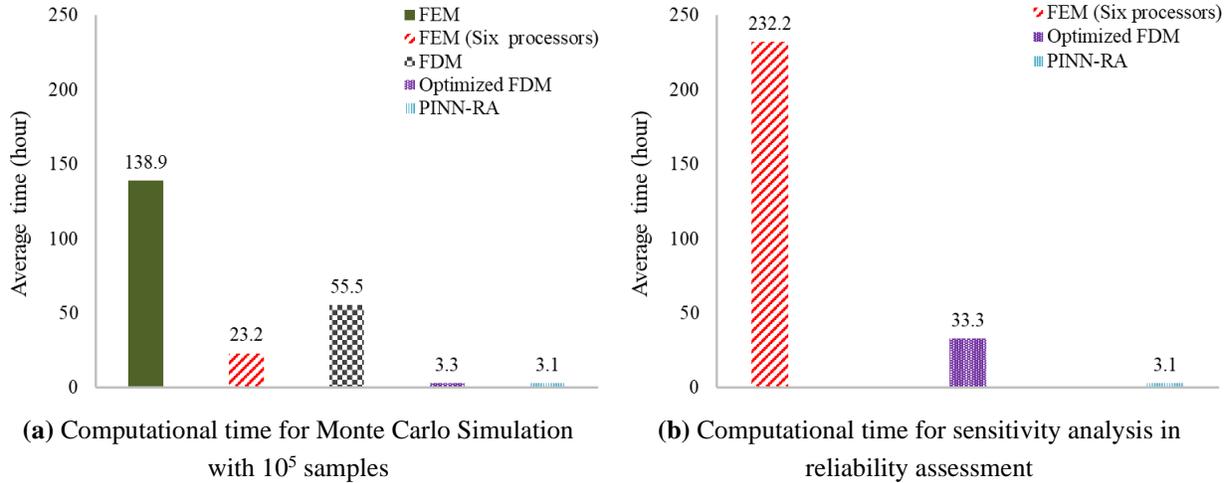

**Fig. 11.** Computational time comparison for reliability analysis using Monte Carlo Simulation between the PINN-RA model for the biaxial stress state, FEM, and FDM

## 6. Summary and Conclusion

This study presents a newly proposed Physics-Informed Neural Network for Reliability Analysis (PINN-RA) approach for pipelines subjected permanent ground movement. It is built based on an extended PINN-based surrogate model, which is developed to account for bi-axial material plasticity and operational loads such as internal pressure and temperature change. To allow its use for reliability analysis considering uncertain variables, particularly those related to soil properties, the PINN model is further extended to solve a parametric differential equation system, governing pipelines embedded in soil with different soil properties. In the PINN-RA approach, random variables, including ground movement magnitude and uncertain soil properties, are introduced as inputs to the neural network. Monte Carlo simulation is then integrated into the PINN-based surrogate model to calculate the failure probability, where the failure is defined as the strain exceedance. Key findings of this study based on the use of PINN-RA include:

- The PINN-RA-based model effectively quantified the Probability of Failure (PoF) for the pipeline subjected ground movements.
- The biaxial stress state under an operating pressure of 10.26 MPa exhibited lower variability in tensile strain demand compared to the uniaxial stress state, with a mean of 1.92 % and a standard deviation of 0.17 % against 2.25 % and 0.22 %, respectively, contributing to enhanced structural reliability.

- The compressive strain demand demonstrated a narrower spread than the tensile strain demand, with mean values of −0.50 % and −0.88 % and standard deviations of 0.03% and 0.06% under the uniaxial and biaxial stress states, respectively, indicating that tensile strain is more susceptible to variations in the input parameters.
- Sensitivity analysis identified ground movement magnitude (PGD magnitude) as the most critical factor influencing the probability of failure, with sensitivity index values significantly exceeding those of other parameters. Among the soil properties, friction angle and burial depth showed a more pronounced impact on reliability. In contrast, soil cohesion and unit weight exhibited lower sensitivity, indicating a comparatively smaller influence on the reliability assessment.
- Computational efficiency analysis demonstrated the advantages of the PINN-RA over traditional methods. Evaluating $10^5$ samples required 138.9 hours with FEM and 55.5 hours with FDM, while parallelizing FEM with six processors and optimizing FDM reduced these to 23.2 and 3.3 hours, respectively. In contrast, training the PINN-RA model took only 3.1 hours, with sample evaluation using the PINN-based surrogate model remaining nearly instantaneous and independent of sample size. For sensitivity analysis on 10 cases, each with $10^5$ samples, the parallelized FEM model required 232.2 hours, the optimized FDM model 33.3 hours, whereas the PINN-RA maintained its efficiency with a fixed 3.1-hour training time.